\input harvmac
\overfullrule=0pt

%
\def\sqr#1#2{{\vbox{\hrule height.#2pt\hbox{\vrule width
.#2pt height#1pt \kern#1pt\vrule width.#2pt}\hrule height.#2pt}}}

\def\half{{\textstyle{1\over 2}}}

\def\half{{\textstyle{1\over 2}}}

\Title{ \vbox{\baselineskip12pt}}
{\vbox{\centerline{Correlation Functions in }
\bigskip
\centerline{Berkovits' Pure Spinor Formulation}}}
\smallskip
\centerline{Gautam Trivedi
\foot{gautamt@physics.unc.edu}
}
\smallskip
\centerline{\it Department of Physics}
\centerline{\it
University of North Carolina, Chapel Hill, NC 27599-3255}
\bigskip
\smallskip

\noindent We use Berkovits' pure spinor quantization to compute
various three-point tree correlation functions in  position-space for the Type IIB superstring. We solve the constraint equations for the vertex operators and obtain explicit expressions for the graviton and axion components of the vertex operators. Using these operators we compute tree level correlation functions in flat space and discuss their extension to the $AdS_{5} \times S^{5}$ background.
\Date{}

\nref\btenflat{N. Berkovits, ``Super-Poincare Covariant
Quantization of the Superstring'', JHEP 0004: 018, 2000;
hep-th/0001035.}

\nref\btenflatamp{N. Berkovits, B.C. Vallilo, ``Consistency of
Super-Poincare Covariant Superstring Tree Amplitudes'', JHEP 0007:
015, 2000; hep-th/0004171.}

\nref\bcoh{N. Berkovits, ``Cohomology in the Pure Spinor Formalism
for the Superstring'', JHEP 0009: 046, 2000; hep-th/0006003.}

\nref\bannarb{N. Berkovits, ``Covariant Quantization of the
Superstring'', Presented at Strings '00, Ann Arbor, Michigan, Int.
J. Mod. Phys. A16:801, 2001; hep-th/0008145.}

\nref\bvoadsfive{N. Berkovits and O. Chandia, ``Superstring Vertex
Operators in an $AdS_5\times S^5$ Background'', Nucl. Phys. B596:
185, 2001; hep-th/0009168.}

\nref\brns{N. Berkovits, ``Relating the RNS and Pure Spinor Formalisms for the Superstring'', JHEP 0108: 026, 2001;
hep-th/0104247.}

\nref\blicoh{N. Berkovits and O. Chandia, ``Lorentz Invariance of
the Pure Spinor BRST Cohomology for the Superstring'', Phys. Lett.
B514: 394, 2001; hep-th/0105149.}

\nref\berkhow{N. Berkovits, P. Howe, ``Ten-Dimensional Supergravity Constraints from the Pure Spinor Formalism for the Superstring''; hep-th/0112160.}

\nref\bcmvo{N. Berkovits, O. Chandia, ``Massive Superstring Vertex Operator in 
 D=10 Superspace'', hep-th/0204121.}

\nref\berkPP{N. Berkovits, ``Conformal Field Theory for the Superstring in a 
Ramond-Ramond Plane Wave Background''; hep-th/0203248.}

\newsec{Introduction}

In this paper we compute explicitly several tree level string
correlation functions for the Type IIB superstring using the pure spinor formulation developed by Berkovits {\it et al.} \refs{\btenflat-\bcmvo}. This quantization gives us tools to evaluate string correlation functions in a manifestly
supersymmetric and covariant manner. The formalism uses the usual
ten-dimensional superspace coordinates
$x^{m},\theta^{\alpha},\bar\theta^{\bar\alpha}$ and introduces new
worldsheet bosonic fields
$\lambda^{\alpha},\bar\lambda^{\bar\alpha}$ which are spacetime spinors and satisfy the pure-spinor condition $\lambda\gamma\lambda=0$. It also provides a
nilpotent BRST charge $Q$, a Virasoro current with vanishing
conformal anomaly and a ghost current. Physical vertex operators for massless pields have conformal weight zero and are states of ghost number 1 in the cohomology of $Q$. Recently this formalism has been used to obtain vertex operators 
for some of the massive fields of the open superstring \refs{\bcmvo}. It has also been applied to construct a worldsheet action for the superstring in a Ramond-Ramond plane wave background \refs{\berkPP}. 

In sect. 2 of this paper we review the main components of the pure
spinor formalism. We will explicitly write down the vertex
operators for the physical states using their constraint
equations. In sect. 3 we use the  vertex operators and the prescription 
for integration over the zero modes of $\theta$s and $\lambda$s \refs{\btenflat} to compute several flat space string correlation functions in position space. Finally, in sect.4 we discuss an extension of these calculations to the $AdS_{5}\times S^{5}$ background. We find the string amplitudes calculated using this proceedure to be equal to the field theory expressions, in accordance with the non-renormalisation theorems for the super-symmetric three-point functions. We expect $\alpha^{\prime}$ corrections to first appear in the four point string tree amplitudes. Berkovits and Vallilo \refs{\btenflatamp} have given a formal proof of the equivalence of the superstring amplitudes in their formulation with the Ramond-Neveu-Schwarz (RNS) quantization, at least in flat space. We present an explicit calculation in order to elucidate the computation of flat space correlation functions in the new quantization and to facilitate the extension to curved backgrounds.

\newsec{Review of Components}

We start by listing the worldsheet fields in the formulation.
There are the usual ten-dimensional superspace coordinates
$x^{m},\theta^{\alpha},\bar\theta^{\bar\alpha}$ where $1 \leq
\alpha, \bar\alpha \leq 16$. In addition to these there are
worldsheet bosons $\lambda^{\alpha}, \bar\lambda^{\bar\alpha}$
which satisfy the following condition \eqn\pscon{
\lambda^{\alpha}\gamma^{m}_{\alpha\beta}\lambda^{\beta}=0, \, \quad \bar\lambda^{\bar\alpha}\gamma^{m}_{\bar\alpha\bar\beta}\bar\lambda^{\bar\beta}=0, \qquad
0\leq m\leq 9.} Here $\gamma^{m}$ are the off-diagonal components
of the 32$\times$32 ten-dimensional $\gamma$ matrices in a Weyl
representation (see Appendix A).

\noindent
The massless vertex operator expanded in powers of $\theta$ and
$\bar\theta$ is \eqn\fullvxop{\eqalign { V(x,\theta,\bar\theta)=
\lambda^{\alpha}\bar\lambda^{\bar\alpha}[
&h_{mn}\gamma^{m}_{\alpha\beta}\gamma^{n}_{\bar\alpha\bar\beta}\theta^{\beta}\bar\theta^{\bar\beta}\cr
&\bar\psi^{\bar\beta}_{m}\gamma^{m}_{\alpha\beta}\gamma_{qrs\bar\alpha\bar\beta}\gamma^{qrs}_{\bar\rho\bar\sigma}\theta^{\beta}
\bar\theta^{\bar\rho}\bar\theta^{\bar\sigma}
+\psi^{\beta}_{m}\gamma_{qrs\alpha\beta}\gamma^{qrs}_{\rho\sigma}\gamma^{m}_{\bar\alpha\bar\beta}\theta^{\rho}\theta^{\sigma}
\bar\theta^{\bar\beta}\cr
&+F^{\beta\bar\beta}\gamma_{mnp\alpha\beta}\gamma^{mnp}_{\rho\sigma}\gamma_{qrs\bar\alpha\bar\beta}\gamma^{qrs}_{\bar\rho\bar\sigma}
\theta^{\rho}\theta^{\sigma}\bar\theta^{\bar\rho}\bar\theta^{\bar\sigma}+\cdots] \, .
}} Here $F^{\beta\bar\beta}$ corresponds to the Ramond-Ramond
field strengths and can be expanded as
\eqn\Fexp{F^{\beta\bar\beta}=C_{m}\gamma^{m\beta\bar\beta}+H_{mnp}\gamma^{mnp\beta\bar\beta}
+F_{mnpqr}\gamma^{mnpqr\beta\bar\beta}.} where $C_{m} \equiv
\partial_{m}\Phi$, for example, corresponds to the field strength for the axion and ($\cdots$) 
correspond  to auxilliary terms with higher powers of $\theta(\bar\theta)$.
\noindent
One now defines the BRST operator as follows \refs{\btenflat}
\eqn\flatbrst{\eqalign{ &Q=\oint dz \lambda^{\alpha} d_{\alpha}\cr
&{\rm where}\cr &d_{\alpha}=p_{\alpha}-{1\over
2}\gamma^{m}_{\alpha\beta}\theta^{\beta}\partial x_{m}+{1\over
8}\gamma^{m}_{\alpha\beta}\gamma_{m\rho\sigma}\theta^{\beta}
\theta^{\rho}\partial\theta^{\sigma} \, .\cr}} where $p_{\alpha}$ are
the conjugate momenta for the $\theta^{\alpha}$s, with similar expressions for $\bar{Q}$. The constraint
equations for the vertex operator are 
\eqn\constraint{\eqalign{ &[Q,V(z,\bar{z})]=0 \, , \quad [\bar{Q},V(z,\bar{z})]=0 \, .\cr }}
In this paper we consider the vertex operators for the graviton
and the axion explicitly. 
Equations \constraint\ imply that the simple poles for the OPEs between $Q,\bar{Q}$ and $V(z,\bar{z})$ vanish. This leads to following differential equations for $V(z,\bar{z})$.
\eqn\constdiff{\eqalign{
&\lambda^{\alpha}\bar\lambda^{\bar\alpha}D_{\alpha}\bar{D}_{\bar\alpha}V=0 \cr
&D_{\alpha}={\partial \over \partial\theta^{\alpha}} + \gamma^{m}_{\alpha\beta}\theta^{\beta}{\partial \over \partial x^m} \quad , \quad \bar{D}_{\bar\alpha}={\partial \over \partial\bar\theta^{\bar\alpha}} + \gamma^{m}_{\bar\alpha\bar\beta}\bar\theta^{\bar\beta}{\partial \over \partial x^m} \, . \cr
}}
Using \constdiff\ we see that the terms with odd powers of $\theta(\bar\theta)$ are related to each other and so are the even powers. Next, we pick out the graviton and the axion vertex operators.
\eqn\grvop{ \eqalign{
V_{graviton}=\lambda^{\alpha}\bar{\lambda}^{\bar\alpha}
&[h_{mn}(x)\gamma^{m}_{\alpha\beta}\gamma^{n}_{\bar\alpha\bar\beta}
\theta^{\beta}\bar\theta^{\bar\beta}\cr &-\half\partial_{\bar
m}h_{m\bar n}(x)\gamma^{m}_{\alpha\beta}\gamma_{\bar
p\bar\alpha\bar\beta}\gamma^{\bar m\bar n\bar
p}_{\bar\rho\bar\sigma}
\theta^{\beta}\bar\theta^{\bar\beta}\bar\theta^{\bar\rho}\bar\theta^{\bar\sigma}\cr
&-\half\partial_mh_{n\bar
m}(x)\gamma_{p\alpha\beta}\gamma^{mnp}_{\rho\sigma}\gamma^{\bar
m}_{\bar\alpha\bar\beta}\theta^{\beta}\theta^{\rho}\theta^{\sigma}\bar\theta^{\bar\beta}\cr
&+{\textstyle{1\over4}}\partial_{m}\partial_{\bar m}h_{n\bar
n}(x)\gamma_{p\alpha\beta}\gamma^{mnp}_{\rho\sigma}\gamma_{\bar
p\bar\alpha\bar\beta}\gamma^{\bar m\bar n\bar
p}_{\bar\rho\bar\sigma}\theta^{\beta}\theta^{\rho}\theta^{\sigma}\bar\theta^{\bar\beta}\bar\theta^{\bar\rho}\bar\theta^{\bar\sigma}\cr
&+\cdots]. }} where $h_{mn}$ is symmetric and traceless and
satisfies $\partial^{p}\partial_{p}h_{mn}=0$ and $
\partial^{m}h_{mn}=0$ and ($\cdots$) correspond to terms that have
higher powers of $\theta (\bar\theta)$  and do not contribute to
tree amplitudes.
\noindent
The vertex operator for the axion is \eqn\axvop{
V_{axion}=\lambda^{\alpha}\bar\lambda^{\bar\alpha}[\partial_q\Phi(x)\gamma^{q\kappa\bar\kappa}\gamma_{mnp\alpha\kappa}\gamma^{mnp}_{\rho\sigma}\gamma_{\bar
m\bar n\bar p\bar\alpha\bar\kappa}\gamma^{\bar m\bar n\bar
p}_{\bar\rho\bar\sigma}\theta^{\rho}\theta^\sigma\bar\theta^{\bar\rho}\bar\theta^{\bar\sigma}+\cdots].
} With results from the appendix this can be written in a more convenient form as \eqn\axvopsim{
V_{axion}={1 \over 16}\lambda^{\alpha}\bar\lambda^{\bar\alpha}[\partial_q\Phi(x)\gamma^{q\kappa\bar\kappa}\gamma_{m\alpha\rho}\gamma^{m}_{\kappa\sigma}
\gamma_{\bar m\bar\alpha\bar\rho}\gamma^{\bar
m}_{\bar\kappa\bar\sigma}
\theta^{\rho}\theta^{\sigma}\bar\theta^{\bar\rho}\bar\theta^{\bar\sigma}+\cdots].
}
\noindent
In the next section we use these vertex operators to calculate
the three-graviton and the two-axion one-graviton correlation
functions in flat space.

\newsec{Correlation functions}
\subsec{Three-graviton tree amplitude in flat space}
\noindent
The three-graviton amplitude is as follows \eqn\gramp{ \eqalign{
A_{ggg}=&<0|V_{graviton}(z_{1},\bar z_{1})V_{graviton}(z_{2},\bar
z_{2})V_{graviton}(z_{3},\bar z_{3})|0>.\cr }}

\noindent
Using \grvop\ we will find two different types of terms that will contribute to the amplitude. These will have the form $ h_{mn}h_{pq}\partial_{r}\partial_{s}h_{tu}$ and $  h_{mn}\partial_{r}h_{pq}\partial_{s}h_{tu}$. There will be three terms of the first kind and six of the second. We begin by looking at a term of the first kind.

\eqn\termOne{ \eqalign{ Term1\equiv
&{1\over4}<\lambda^{\alpha}\lambda^{\beta}\lambda^{\gamma}\bar\lambda^{\bar\alpha}\bar\lambda^{\bar\beta}\bar\lambda^{\bar\gamma}h_{mn}(x)h_{pq}(x)\partial_{r}\partial_{s}h_{tu}(x)\cr
&\times\gamma^{m}_{\alpha\rho}\gamma^{p}_{\beta\sigma}\gamma_{w\gamma\delta}\gamma^{rtw}_{\kappa\tau}\gamma^{n}_{\bar\alpha\bar\rho}\gamma^{q}_{\bar\beta\bar\sigma}\gamma_{v\bar\gamma\bar\delta}\gamma^{suv}_{\bar\kappa\bar\tau}\cr
&\times\theta^{\rho}\theta^{\sigma}\theta^{\delta}\theta^{\kappa}\theta^{\tau}
\bar\theta^{\bar\rho}\bar\theta^{\bar\sigma}\bar\theta^{\bar\delta}\bar\theta^{\bar\kappa}\bar\theta^{\bar\tau}>\cr
=&{1\over4}<h_{mn}(x)h_{pq}(x)\partial_{r}\partial_{s}h_{tu}(x)>_{X}\cr
&<\lambda^{\alpha}\lambda^{\beta}\lambda^{\gamma}\bar\lambda^{\bar\alpha}\bar\lambda^{\bar\beta}\bar\lambda^{\bar\gamma}
\gamma^{m}_{\alpha\rho}\gamma^{p}_{\beta\sigma}\gamma_{w\gamma\delta}\gamma^{rtw}_{\kappa\tau}\gamma^{n}_{\bar\alpha\bar\rho}\gamma^{q}_{\bar\beta\bar\sigma}\gamma_{v\bar\gamma\bar\delta}\gamma^{suv}_{\bar\kappa\bar\tau}\cr
&\times\theta^{\rho}\theta^{\sigma}\theta^{\delta}\theta^{\kappa}\theta^{\tau}
\bar\theta^{\bar\rho}\bar\theta^{\bar\sigma}\bar\theta^{\bar\delta}\bar\theta^{\bar\kappa}\bar\theta^{\bar\tau}>\, .\cr
}}
Since $\lambda$ and $\theta$ have conformal weight zero, only their zero modes will survive the bracket. In accordance with Berkovits' prescription \refs{\btenflat-\bcoh}, we are keeping only terms containing five $\theta$s and three $\lambda$s. These are the only ones that will contribute to the amplitude because there is only one state $(\lambda\gamma\theta)(\lambda\gamma\theta)(\lambda\gamma\theta)(\theta\gamma\theta)|0>$ in the cohomology of $Q$ with ghost  number three. We can show that
\eqn\presc{\eqalign{
<\lambda^{\alpha}\lambda^{\beta}\lambda^{\gamma}\theta^{\rho}\theta^{\sigma}\theta^{\tau}
\theta^{\omega}\theta^{\kappa}>=&T^{\alpha\beta\gamma}_{\alpha_{1}\beta_{1}\gamma_{1}}
\gamma_{q}^{\alpha_{1}[\rho}\gamma_{r}^{\beta_{1}\sigma}\gamma_{s}^{\gamma_{1}\tau}\gamma^{qrs\omega\kappa]}\cr}}
where [$\, \, $] stands for antisymmetrization over the indices $\rho ,\sigma ,\tau
,\omega$ and $ \kappa$, with no overall normalization factor. One can obtain a similar expression for
$<\bar\lambda^{\bar\alpha}\bar\lambda^{\bar\beta}\bar\lambda^{\bar\gamma}\bar\theta^{\bar\rho}
\bar\theta^{\bar\sigma}\bar\theta^{\bar\tau}\bar\theta^{\bar\omega}\bar\theta^{\bar\kappa}>$.

\noindent Here
$T^{\alpha\beta\gamma}_{\alpha_{1}\beta_{1}\gamma_{1}}$ is defined \refs{\btenflat}
as \eqn\T{T^{\alpha\beta\gamma}_{\alpha_{1}\beta_{1}\gamma_{1}}=
{N \over
4032}[\delta^{(\alpha}_{\alpha_{1}}\delta^{\beta}_{\beta_{1}}\delta^{\gamma)}_{\gamma_{1}}
-{1 \over
40}\gamma^{(\alpha\beta}_{m}\delta^{\gamma)}_{(\alpha_{1}}\gamma^{m}_{\beta_{1}\gamma_{1})}] \, .
}
\noindent
The brackets $( \, \,)$ corresponds to symmetrization over the enclosed indices with no overall normalization, and $T^{\alpha\beta\gamma}_{\alpha\beta\gamma}=N$.
\noindent Using \presc\ we evaluate the following useful terms.
\eqn\norm{\eqalign{
<\lambda^{\alpha}\lambda^{\beta}\lambda^{\gamma}
\gamma_{m\alpha\rho}\gamma_{n\beta\tau}\gamma_{p\gamma\sigma}
\gamma^{mnp}_{\kappa\delta}\theta^{\rho}\theta^{\tau}\theta^{\sigma}
\theta^{\kappa}\theta^{\delta}>\, \, =\,\, &2880.\cr
<\lambda^{\alpha}\lambda^{\beta}\lambda^{\gamma}
\gamma^{t}_{\alpha\rho}\gamma_{n\beta\tau}\gamma_{p\gamma\sigma}
\gamma^{mnp}_{\kappa\delta}\theta^{\rho}\theta^{\tau}\theta^{\sigma}
\theta^{\kappa}\theta^{\delta}>\,\, =\,\, &288\eta^{mt}.\cr
<\lambda^{\alpha}\lambda^{\beta}\lambda^{\gamma}
\gamma^{t}_{\alpha\rho}\gamma^{u}_{\beta\tau}\gamma_{p\gamma\sigma}
\gamma^{mnp}_{\kappa\delta}\theta^{\rho}\theta^{\tau}\theta^{\sigma}
\theta^{\kappa}\theta^{\delta}>\,\, =\,\, &32\eta^{t[m}\eta^{n]u}.\cr
<\lambda^{\alpha}\lambda^{\beta}\lambda^{\gamma}
\gamma^{t}_{\alpha\rho}\gamma^{u}_{\beta\tau}\gamma^{v}_{\gamma\sigma}
\gamma^{mnp}_{\kappa\delta}\theta^{\rho}\theta^{\tau}\theta^{\sigma}
\theta^{\kappa}\theta^{\delta}>\,\,=\,\, &4(\eta^{tm}\eta^{nu}\eta^{pv}+\eta^{tn}\eta^{up}\eta^{vm}+\eta^{tp}\eta^{um}\eta^{vn}\cr
&-\eta^{tn}\eta^{um}\eta^{vp}-\eta^{tp}\eta^{un}\eta^{vm}-\eta^{tm}\eta^{up}\eta^{vn}) \, . }} 
\noindent
Note that the normalization used in \norm\ corresponds to  $N= {1 \over 304}$, which follows from a long but straightforward calculation of traces over various combinations of $\gamma$-matrices. Using the third term from \norm\ in \termOne\ we find 
\eqn\termOnetwo{\eqalign{
Term1&=256(\eta^{mr}\eta^{tp}\eta^{ns}\eta^{uq}+\eta^{mt}\eta^{rp}
\eta^{nu}\eta^{sq}-\eta^{mt}\eta^{rp}\eta^{ns}\eta^{uq}-\eta^{mr}\eta^{tp}\eta^{nu}\eta^{sq})\cr
&\times<h_{mn}(x)h_{pq}(x)\partial_{r}\partial_{s}h_{tu}(x)>_{X}\cr
&=256<2h_{mn}(x)h_{pq}(x)\partial^{m}\partial^{n}h^{
pq}(x)+2h_{mn}(x)\partial^{n}h_{pq}(x)\partial^{p}h^{mq}(x)>_{X}}}
Similarly we now evaluate a term of the second kind.
\eqn\termTwo{\eqalign{ Term2\equiv
&-{1\over4}<\lambda^{\alpha}\lambda^{\beta}\lambda^{\gamma}\bar\lambda^{\bar\alpha}
\bar\lambda^{\bar\beta}\bar\lambda^{\bar\gamma}h_{mn}(x)\partial_{p}h_{qr}(x)\partial_{s}h_{tu}(x)\cr
&\times
\gamma^{m}_{\alpha\rho}\gamma^{n}_{\bar\alpha\bar\rho}\gamma^{q}_{\beta\sigma}\gamma_{v\bar\beta\bar\tau}
\gamma^{prv}_{\bar\kappa\bar\delta}\gamma_{w\gamma\tau}\gamma^{stw}_{\kappa\delta}\gamma^{u}_{\bar\gamma\bar\sigma}\cr
&\times
\theta^{\rho}\theta^{\sigma}\theta^{\tau}\theta^{\kappa}\theta^{\delta}
\bar\theta^{\bar\rho}\bar\theta^{\bar\sigma}\bar\theta^{\bar\tau}\bar\theta^{\bar\kappa}\bar\theta^{\bar\delta}>\cr
=&-256(\eta^{ms}\eta^{tq}\eta^{np}\eta^{ru}+\eta^{mt}\eta^{sq}\eta^{nr}\eta^{pu}-\eta^{mt}\eta^{sq}\eta^{np}\eta^{ru}-\eta^{ms}\eta^{tq}\eta^{nr}\eta^{pu})\cr
&\times<h_{mn}(x)\partial_{p}h_{qr}(x)\partial_{s}h_{tu}(x)>_{X}\cr
=&256<h_{mn}(x)h_{pq}(x)\partial^{m}\partial^{n}h^{pq}(x)
+3h_{mn}(x)\partial^{n}h_{pq}(x)\partial^{p}h^{mq}(x)>_{X}}}
Combining \termOnetwo\ and \termTwo\ we compute the amplitude to be
\eqn\gggamp{\eqalign{ A_{ggg}=&3Term1+6Term2\cr
=&3072 \int d^{10}x \,h_{mn}(x)h_{pq}(x)\partial^{m}\partial^{n}h^{pq}(x)+2h_{mn}(x)\partial^{n}h_{pq}(x)\partial^{p}h^{mq}(x) \, .}}
This is proportional to the field theory result for the three-graviton amplitude in a flat background on shell in the $\partial^{m}h_{mn}=0$ gauge.

\subsec{Two-axion one-graviton tree amplitude in flat space}

Using \grvop\ and \axvopsim\ we can compute the 2-axion, 1-graviton scattering amplitude as 

\eqn\axaxgr{ \eqalign{ A_{aag}=&<V_{graviton}(z_{1},\bar
z_{1})V_{axion}(z_{2},\bar z_{2})V_{axion}(z_{3},\bar z_{3})>\cr
A_{aag}=&{1 \over 256}<\lambda^\alpha\lambda^\beta\lambda^\gamma\bar\lambda^{\bar\alpha}\bar\lambda^{\bar\beta}\bar\lambda^{\bar\gamma}h_{mn}(x)\partial_{p}\Phi(x)\partial_{q}\Phi(x)\gamma^{m}_{\alpha\rho}\gamma^{n}_{\bar\alpha\bar\rho}\cr
&\gamma^{p\kappa\bar\kappa}\gamma_{r\beta\sigma}\gamma^{r}_{\kappa\tau}\gamma_{s\bar\beta\bar\sigma}\gamma^{s}_{\bar\kappa\bar\tau}\gamma^{q\eta\bar\eta}\gamma_{t\gamma\xi}\gamma^{t}_{\eta\pi}\gamma_{u\bar\gamma\bar\xi}\gamma^{u}_{\bar\eta\bar\pi}\cr
&\theta^{\rho}\theta^{\sigma}\theta^{\tau}\theta^{\xi}\theta^{\pi}\bar\theta^{\bar\rho}\bar\theta^{\bar\sigma}\bar\theta^{\bar\tau}\bar\theta^{\bar\xi}\bar\theta^{\bar\pi}>\cr
=&{1 \over 256}<h_{mn}(x)\partial_{p}\Phi(x)\partial_{q}\Phi(x)>_{X}\cr
&<\lambda^\alpha\lambda^\beta\lambda^\gamma\bar\lambda^{\bar\alpha}\bar\lambda^{\bar\beta}\bar\lambda^{\bar\gamma}\gamma^{m}_{\alpha\rho}\gamma^{n}_{\bar\alpha\bar\rho}
\gamma^{p\kappa\bar\kappa}\gamma_{r\beta\sigma}\gamma^{r}_{\kappa\tau}\gamma_{s\bar\beta\bar\sigma}\gamma^{s}_{\bar\kappa\bar\tau}\gamma^{q\eta\bar\eta}\gamma_{t\gamma\xi}\gamma^{t}_{\eta\pi}\gamma_{u\bar\gamma\bar\xi}\gamma^{u}_{\bar\eta\bar\pi}\cr
&\theta^{\rho}\theta^{\sigma}\theta^{\tau}\theta^{\xi}\theta^{\pi}\bar\theta^{\bar\rho}\bar\theta^{\bar\sigma}\bar\theta^{\bar\tau}\bar\theta^{\bar\xi}\bar\theta^{\bar\pi}> \, .
}}
\noindent
We now define \eqn\Aax{\eqalign{
A^{mnpq}=&<\lambda^\alpha\lambda^\beta\lambda^\gamma\bar\lambda^{\bar\alpha}\bar\lambda^{\bar\beta}\bar\lambda^{\bar\gamma}\gamma^{m}_{\alpha\rho}\gamma^{n}_{\bar\alpha\bar\rho}
\gamma^{p\kappa\bar\kappa}\gamma_{r\beta\sigma}\gamma^{r}_{\kappa\tau}\gamma_{s\bar\beta\bar\sigma}\gamma^{s}_{\bar\kappa\bar\tau}\gamma^{q\eta\bar\eta}\gamma_{t\gamma\xi}\gamma^{t}_{\eta\pi}\gamma_{u\bar\gamma\bar\xi}\gamma^{u}_{\bar\eta\bar\pi}\cr
&\theta^{\rho}\theta^{\sigma}\theta^{\tau}\theta^{\xi}\theta^{\pi}\bar\theta^{\bar\rho}\bar\theta^{\bar\sigma}\bar\theta^{\bar\tau}\bar\theta^{\bar\xi}\bar\theta^{\bar\pi}>\cr
}}
We see that $A^{mnpq}=A^{mnqp}$. Also, since $h_{mn}$ is traceless, the only
component of $A^{mnpq}$ that suvives must be
$C\delta^{m(p}\delta^{q)n}$, where $C$ is some constant. It follows that

\eqn\axaxgramp{\eqalign{
A_{aag}=&C \int d^{10}x \, h_{mn}(x)\partial^{m}\Phi(x)\partial^{n}\Phi(x)\cr
}}
To calculate $C$ requires another long but
straightforward trace calculation.

\newsec{Graviton two-axion amplitude in $AdS_5\times S^5$}
In this section we will discuss correlation functions on the $AdS_{5}\times S^{5}$ background. We will consider the two-axion one-graviton amplitude $A_{aag}$. For $AdS_{5}\times S^{5}$, we introduce the curved space gamma matrices which satisfy $\gamma^{m\alpha\beta}\gamma^{n}_{\beta\gamma}+\gamma^{n\alpha\beta}\gamma^{m}_{\beta\gamma}=2\bar{g}^{mn}\delta^{\alpha}_{\gamma}$, where $\bar{g}^{mn}$ is the $AdS_{5}\times S^{5}$ background metric. Similarly for the barred indices we have $\gamma^{m\bar\alpha\bar\beta}\gamma^{n}_{\bar\beta\bar\gamma}+\gamma^{n\bar\alpha\bar\beta}\gamma^{m}_{\bar\beta\bar\gamma}=2\bar{g}^{mn}\delta^{\bar\alpha}_{\bar\gamma}$. In this case, following \refs{\btenflat}, we can convert between the barred and unbarred spinor indices using $\delta^{\alpha\bar\alpha}=(\gamma^{01234})^{\alpha\bar\alpha}$, {\it i.e.} $G^{\alpha}=\delta^{\alpha\bar\alpha}G_{\bar\alpha}$ and $G^{\bar\alpha}=\delta^{\alpha\bar\alpha}G_{\alpha}$. Here [01234] are the $AdS_{5}$ directions.
Since $\delta^{\alpha\bar\alpha}$ is an orthogonal matrix, contracting over two barred indices is the same as contracting over two unbarred indices, {\it i.e.} $G_{\alpha}G^{\alpha}=G_{\bar\alpha}G^{\bar\alpha}$. 

We now need to compute \axaxgr\ using  the curved space {$\gamma$-matrices \foot{In principle, to establish \axaxgr\ for the curved background, one also has to know the OPEs for $Q$ and $\bar{Q}$. For this particular amplitude however, we do not need the auxilliary terms.}}. Our rule for summing over barred indices implies that the computation is similar to taking the trace flat space, but now the flat metric $\eta^{mn}$ is replaced by $\bar{g}^{mn}$. Furthermore, we can promote the partial derivatives to covariant derivatives since they only act on the scalar axion.
Therefore, we suggest that the two-axion one-graviton string tree amplitude for Type IIB strings on $AdS_{5}\times S^{5}$ can be obtained by covariantizing the flat space result and will have the form
\eqn\aagcrv{
A_{aag}=C^{\prime}\int d^{10}x {\sqrt {\bar g}} \, \,\bar{g}_{mp}\bar{g}_{nq}h^{mn}\bar{D}^{p}\Phi\bar{D}^{q}\Phi \, .
}

\noindent
In order to calculate more general amplitudes on the $AdS_{5}\times S^{5}$ background, one would need to derive the invariant derivatives from the string constraint equations. One would also need to establish the auxilliary terms in the vertex operators, using either the constraint equations or symmetry arguments.

\appendix{A}{Gamma matrices.}

\noindent
This appendix describes the $\gamma$-matrices used in this paper.
In the Weyl representation, the  ten-dimensional gamma matrices are defined as 
\eqn\gamten{
{\gamma^{mA}}_{B}= \left(\matrix{
0 & {\gamma^{m\alpha}}_{\bar{\beta}}\cr
{\gamma^{m}_{\bar{\alpha}}}^{\beta} & 0
}\right)}
where $1 \leq A,B \leq 32$ and $1 \leq \alpha , \bar{\alpha}, \beta , \bar{\beta} \leq 16$. The ten-dimensional charge conjugation matrix is

\eqn\cc{
C^{AB}= \left(\matrix{
0 & C^{{\alpha}\bar{\beta}}\cr
C^{\bar{\alpha}{\beta}} & 0}\right)}
where
\eqn\ccinv{
C^{-1}_{AB}= \left(\matrix{
0 & C^{-1}_{{\alpha}\bar{\beta}}\cr
C^{-1}_{\bar{\alpha}{\beta}} & 0}\right)\, .}
The charge conjugation matrix and its inverse can be used to raise and lower
the indices on the gamma matrices as $\gamma^{mAB} = {\gamma^{mA}}_{D}C^{DB}$ and $\gamma^{m}_{AB} = C^{-1}_{AD}{\gamma^{mD}}_{B}$.
As in \refs{\btenflat} the $\gamma^{m\alpha\beta}$ and $\gamma^{m}_{\alpha\beta}$ are the off-diagonal elements of the ten-dimensional 32 $\times$ 32 Dirac-matrices and satisfy $\gamma^{m}\gamma^{n}+\gamma^{n}\gamma^{m}=2\eta^{mn}$ in flat space.

\noindent
The matrix $\gamma^{\alpha_{1}\alpha_{2}...\alpha_{N}}$ denotes the completely antisymmetric product of $N$ gamma matrices. Specifically,
\eqn\gamcoms{\eqalign{
&\gamma^{m}_{\alpha\beta}=\gamma^{m}_{\beta\alpha}\cr
&\gamma^{mnp}_{\alpha\beta}={1 \over 3!}\gamma^{[m}\gamma^{n}\gamma^{p]}=-\gamma^{mnp}_{\beta\alpha}\cr
&\gamma^{mnpqr}_{\alpha\beta}={1 \over 5!}\gamma^{[m}\gamma^{n}\gamma^{p}\gamma^{q}\gamma^{r]}=\gamma^{mnpqr}_{\beta\alpha} \, .}}
where one can show that $\gamma^{m}$,$\gamma^{mnp}$ and $\gamma^{mnp}$ form a complete basis for expansion of any matrix $A_{\alpha\beta}$.
Some useful results are 
\eqn\gamone{
\gamma_{m(\alpha\beta}\gamma^{m}_{\gamma)\delta}=0}
\eqn\gamthree{
\gamma_{mnp[\alpha\beta}\gamma^{mnp}_{\gamma]\delta}=0}

\noindent 
In order to show \gamthree\ we start by first showing a more general result

\eqn\gamonethree{
\zeta^{\alpha}\gamma^{m}_{\alpha\rho}\theta^{\rho}\eta^{\beta}\gamma_{m\beta\sigma}\theta^{\sigma}={1 \over 16}\zeta^{\alpha}\gamma^{mnp}_{\alpha\beta}\eta^{\beta}\theta^{\rho}\gamma_{mnp\rho\sigma}\theta^{\sigma}
}
\noindent
We note that 
\noindent
\eqn\proofone{\eqalign{
\theta^{\rho}\theta^{\sigma}&=-\theta^{\sigma}\theta^{\rho} \cr
\Rightarrow\theta^{\rho}\theta^{\sigma}&=C^{pqr}\gamma_{pqr}^{\rho\sigma}\cr
\Rightarrow {\rm L.H.S \, \, of \, \,  \gamonethree\ }&=\zeta^{\alpha}\eta^{\beta}\gamma^{m}_{\alpha\rho}\gamma_{m\sigma\beta}C^{pqr}\gamma_{pqr}^{\rho\sigma}\cr
}}

\noindent
Let us now look at the following term:
\eqn\prooftwo{\eqalign{
\gamma^{m}_{\alpha\rho}\gamma_{pqr}^{\rho\sigma}\gamma_{m\sigma\beta}&=-\gamma^{m}_{\beta\rho}\gamma_{pqr}^{\rho\sigma}\gamma_{m\sigma\alpha}\cr
\Rightarrow\gamma^{m}_{\alpha\rho}\gamma_{pqr}^{\rho\sigma}\gamma_{m\sigma\beta}&= A_{pqr}^{\ell mn}\gamma_{\ell mn\alpha\beta}
}}

\noindent
One can now evaluate $A_{pqr}^{stu}$ by tracing over the two side of \prooftwo using a $\gamma^{stu\alpha\beta}$. The result one obtains is the following:
\eqn\proofthree{\eqalign{
A_{pqr}^{stu}=-\delta^{s}_{[p}\delta^{t}_{q}\delta^{u}_{r]}\cr}}
Similarly one can show that
\eqn\prooffour{\eqalign{
C^{pqr}={-1 \over 96}\gamma^{pqr}_{\rho\sigma}\theta^{\rho}\theta^{\sigma}}}
Substituting \proofthree\ and \prooffour\ in \proofone we get
\eqn\QED{\eqalign{
{\rm L.H.S \, \, of \, \, \gamonethree\ }=&{1 \over 16}\zeta^{\alpha}\eta^{\beta}\gamma_{pqr\rho\sigma}\theta^{\rho}\theta^{\sigma}\gamma^{pqr}_{\alpha\beta}\cr
=&{\rm R.H.S \, \, of \, \, \gamonethree\ } }}

\noindent
\gamthree\ then follows from \gamonethree\ if we choose $\eta=\theta$ and the use the fact that $\gamma^{m}_{\alpha\beta}=\gamma^{m}_{\beta\alpha}$ .

\vskip50pt {\bf Acknowledgements:} We thank Nathan Berkovits and
Louise Dolan for conversations. GT is partially supported by the
U.S. Department of Energy, Grant No. DE-FG02-97ER-41036/Task A.

\listrefs

\bye